\documentclass[amsmath,aps,showpacs,a4paper,10pt,superscriptaddress]{revtex4}

 \usepackage{epsf}
 \usepackage{graphicx}    

 \allowdisplaybreaks[4]   

\begin{document}

 \newcommand{\be}[1]{\begin{equation}\label{#1}}
 \newcommand{\ee}{\end{equation}}
 \newcommand{\bea}{\begin{eqnarray}}
 \newcommand{\eea}{\end{eqnarray}}
 \def\disp{\displaystyle}

 \def\gsim{ \lower .75ex \hbox{$\sim$} \llap{\raise .27ex \hbox{$>$}} }
 \def\lsim{ \lower .75ex \hbox{$\sim$} \llap{\raise .27ex \hbox{$<$}} }

 \begin{titlepage}

 \begin{flushright}
 arXiv:1602.07189
 \end{flushright}

 \title{\Large \bf New Generalizations of
 Cosmography Inspired by~the~Pad\'e~Approximant}

 \author{Ya-Nan~Zhou\,}
 \email[\,email address:\ ]{nanxiaonan89612@163.com}
 \affiliation{School of Physics,
 Beijing Institute of Technology, Beijing 100081, China}

 \author{De-Zi~Liu\,}
 \affiliation{Department of Astronomy, Peking University,
 Beijing 100871, China}

 \author{Xiao-Bo~Zou\,}
 \affiliation{School of Physics,
 Beijing Institute of Technology, Beijing 100081, China}

 \author{Hao~Wei\,}
 \thanks{\,Corresponding author}
 \email[\,email address:\ ]{haowei@bit.edu.cn}
 \affiliation{School of Physics,
 Beijing Institute of Technology, Beijing 100081, China}

 \begin{abstract}\vspace{1cm}
 \centerline{\bf ABSTRACT}\vspace{2mm}
 The current accelerated expansion of the universe has been one
 of the most important fields in physics and astronomy since
 1998. Many cosmological models have been proposed in the
 literature to explain this mysterious phenomenon. Since the
 nature and cause of the cosmic acceleration are still unknown,
 model-independent approaches to study the evolution of
 the universe are welcome. One of the powerful
 model-independent approaches is the so-called cosmography.
 It only relies on the cosmological principle, without
 postulating any underlying theoretical model. However, there
 are several shortcomings in the usual cosmography. For
 instance, it is plagued with the problem of divergence (or an
 unacceptably large error), and it fails to predict the future
 evolution of the universe. In the present work, we try to
 overcome or at least alleviate these problems, and we~propose
 two new generalizations of cosmography inspired by the Pad\'e
 approximant. One is to directly parameterize the luminosity
 distance based on the Pad\'e approximant, while the other is
 to generalize cosmography with respect to a so-called
 $y_\beta$-shift $y_\beta=z/(1+\beta z)$, which is also
 inspired by the Pad\'e approximant. Then, we confront them
 with the observational data with the help of the Markov chain
 Monte Carlo (MCMC) code {\it emcee}, and find that they
 work fairly well.
 \end{abstract}


 \maketitle

 \end{titlepage}

 \renewcommand{\baselinestretch}{1.0}


\section{Introduction}\label{sec1}

From the observation of distant type
 Ia supernovae (SNIa)~\cite{Riess:1998cb,1999ApJ...517..565P},
 it has been found in 1998 that the universe is
 experiencing an accelerated expansion. This amazing discovery
 was confirmed later by the observations of e.g. cosmic
 microwave background (CMB)~\cite{Page:2003fa,Spergel:2006hy,
 Komatsu:2008hk,Komatsu:2010fb}, and large scale structure
 (LSS)~\cite{Tegmark:2003ud,Seljak:2004xh}. In fact, this
 mysterious phenomenon has been one of the most important
 fields in physics and astronomy.

In the literature (see e.g.~\cite{revofacc} for reviews), there
 are two representative categories of cosmological models
 accounting for the current cosmic acceleration. One is to
 introduce a new component with negative pressure, called
 ``dark energy'', in the right-hand side of the Einstein field
 equation in the framework of general relativity. The other is
 to modify the left-hand side of the Einstein field equation,
 namely to modify general relativity on a cosmological scale
 (known as ``modified gravity theory''). We can constrain dark
 energy models and modified gravity theories by using the
 observational data. However, most of the observational
 constraints are model-dependent in fact. On the other hand,
 both dark energy models and modified gravity theories seem to
 be in agreement with the observational data; the physical
 mechanism to accelerate the cosmic expansion is still unclear
 by now~\cite{revofacc,Bamba:2012cp}. Furthermore, it is argued
 in e.g.~\cite{Kunz:2006ca,Wei:2008vw} that dark energy models
 cannot be distinguished from modified gravity theories
 even by using the observations of both the expansion and the
 growth histories. These confusions suggest that a more
 conservative approach to the problem of the
 cosmic acceleration, relying on as few model-dependent
 quantities as possible, is welcome. Thus,
 various model-independent approaches have been proposed
 in the literature~\cite{revofacc,Bamba:2012cp}. A well-known
 one is the parameterization of equation-of-state parameter
 (EoS), such as $w=w_0+w_1 z$~\cite{Maor:2000jy},
 and $w=w_0+w_a z/(1+z)$~\cite{Chevallier:2000qy}, where
 $z$ is the redshift. Another powerful model-independent
 approach is cosmography~\cite{Weinberg2008,Visser:2004bf,
 Bamba:2012cp,Dunsby:2015ers,Chiba:1998tc,Neben:2012wc,
 Cattoen:2007id,Cattoen:2008th,Aviles:2012ay,
 Capozziello:2008tc,Vitagliano:2009et,Luongo:2011zz,Busti:2015xqa}.
 To the best of our knowledge, it was first discussed by
 Weinberg~\cite{Weinberg2008} and extended by
 Visser~\cite{Visser:2004bf} recently. Using cosmography, one
 can analyze the evolution of the universe without assuming any
 underlying theoretical model. The only necessary assumption of
 cosmography is the cosmological principle, so that the
 spacetime metric is the one of the Friedmann-Robertson-Walker
 (FRW) universe,
 \be{addeq1}
 ds^2=-c^2 dt^2+a^2(t)\left[\frac{dr^2}{1-kr^2}+
 r^2\left(d\theta^2+\sin^2 \theta\,d\phi^2\right)\right]\,,
 \ee
 in terms of the comoving coordinates $(t,\,r,\,\theta,
 \,\phi)$, where $c$ is the speed of light, $a$ is the scale
 factor, and $k>,\,=,\,<0$ corresponds to a spatially close,
 flat, open universe, respectively. Introducing the so-called
 cosmographic parameters, namely the Hubble constant $H_0$,
 the deceleration $q_0$, the jerk $j_0$, the snap $s_0$
 (defined below), one can expand the scale factor $a$ in terms
 of a Taylor series with respect to cosmic
 time~$t$~\cite{Weinberg2008,Visser:2004bf},
 \be{addeq2}
 a(t)=a(t_0)\left[1 + H_0 (t-t_0) - \frac{q_0}{2}
 H_{0}^{2}(t-t_{0})^{2}+\frac{j_{0}}{3!} H_{0}^{3}(t-t_{0})^{3}
 +\frac{s_{0}}{4!} H_{0}^{4} (t-t_{0})^{4}
 +{\cal O}\left((t-t_{0})^5\right)\right]\,,
 \ee
 and also the luminosity distance $d_L$ with respect to
 redshift $z$~\cite{Weinberg2008,Visser:2004bf,
 Bamba:2012cp,Dunsby:2015ers,Chiba:1998tc,Neben:2012wc},
 \bea
 d_L(z)&=&\frac{cz}{H_0}\left\{1+\frac{1}{2}\left(1-q_0\right)z
 -\frac{1}{6}\left[1-q_0-3q_0^2+j_0
 +\frac{kc^2}{H_0^2 a^2(t_0)}\right]z^2\right.\nonumber\\[1mm]
 &&\left.+\frac{1}{24}\left[2-2q_0-15q_0^2-15q_0^3+5j_0+
 10q_0 j_0+s_0+\frac{2kc^2(1+3q_0)}{H_0^2 a^2(t_0)}\right]z^3
 + {\cal O} \left(z^4\right)\right\}\,. \label{addeq3}
 \eea
 So, one can study the universe in a model-independent way
 by using cosmography.

It is easy to see that the key of cosmography is to expand
 the quantities under consideration as a Taylor series with
 respect to redshift $z$. However, it is well known that the
 Taylor series converges only for small $z$ around $0$, and
 it might diverge at high redshift (especially when $z>1$).
 A possible remedy is to replace the redshift $z$ with the
 so-called $y$-shift, $y\equiv z/(1+z)$~\cite{Cattoen:2007id,
 Cattoen:2008th,Vitagliano:2009et}. In this case, $y<1$ holds
 in the whole cosmic history $0\leq z<\infty$, and hence the
 Taylor series with respect to $y$ converges. However, there
 still exist several serious problems in the case
 of $y=z/(1+z)$. The first is that the error of a Taylor
 approximation throwing away the higher order terms will become
 unacceptably large when $y$ is close to~$1$ (say, when $z>9$).
 The second is that the cosmography in terms of $y=z/(1+z)$
 cannot work well in the cosmic future $-1<z<0$. The Taylor
 series with respect to $y=z/(1+z)$ does not converge when
 $y<-1$ (namely $z<-1/2$), and it drastically diverges when
 $z\to -1$ (it is easy to see that $y\to -\infty$ in this
 case). So, this $y$-shift cosmography fails to predict the
 future evolution of the universe. Note that there are other
 $y$-shifts considered in the literature, for instance,
 $y_1\equiv\arctan\left(z/(1+z)\right)$, $y_2\equiv z/(1+z^2)$
 and $y_3\equiv\arctan z$~\cite{Aviles:2012ay}. However, they
 are purely written by hand, without solid motivation. On the
 other hand, $|y|>1$ at suitable redshift $z$ since the
 function $\arctan x\in\left(-\pi/2,\,+\pi/2\right)$, and
 hence the Taylor series does not converge.

In the present work, we try to overcome the problems of
 cosmography mentioned above. We are mainly interested in the
 cosmography of the luminosity distance $d_L$, since it can be
 confronted with the observational data directly. In fact, the
 new generalizations considered in the present work are
 inspired by the so-called Pad\'e approximant. In
 Sec.~\ref{sec2}, we briefly review the key points of the
 Pad\'e approximant, and then we parameterize the luminosity
 distance $d_L$ based on the Pad\'e approximant. We confront
 this parameterization of the luminosity distance $d_L$ with
 the observational data and see whether it works well. Note
 that in this work we use the Markov chain Monte Carlo (MCMC)
 code {\it emcee}~\cite{ForemanMackey:2012ig} in the data
 fitting. In Sec.~\ref{sec3}, inspired by the Pad\'e approximant, we
 propose a new $y_\beta$-shift and then derive the cosmography
 of the luminosity distance $d_L$ by expanding it as a Taylor
 series with respect to the new $y_\beta$-shift. This cosmography is
 completely free from the problems mentioned above. We also confront
 it with the observational data. Finally, some
 brief concluding remarks are given in Sec.~\ref{sec4}.


\section{Pad\'e parameterization of the luminosity distance}\label{sec2}

The so-called Pad\'e approximant can be regarded
 as a generalization of the Taylor series. For any function
 $f(x)$, its Pad\'e approximant of order $(m,\,n)$ is given
 by the rational function~\cite{r3,r4,r5,
 Adachi:2011vu,Gruber:2013wua,Wei:2013jya,Liu:2014vda}
 \be{eq1}
 f(x)=\frac{\alpha_0+\alpha_1 x+\cdots+\alpha_m x^m}{1+
 \beta_1 x+\cdots+\beta_n x^n}\,,
 \ee
 where $m$ and $n$ are both non-negative integers, and
 $\alpha_i$, $\beta_i$ are all constants. Obviously, it
 reduces to the Taylor series when all $\beta_i=0$. Actually
 in mathematics, a Pad\'e approximant is the best approximation
 of a function by a rational function of given order~\cite{r4}.
 In fact, the Pad\'e approximant often gives a better approximation
 of the function than truncating its Taylor series, and it may
 still work where the Taylor series does not converge~\cite{r4}. So,
 considering the Pad\'e approximant in cosmology is well motivated.

Here, we directly parameterize the luminosity distance $d_L$
 based on the Pad\'e approximant,
 \be{eq2}
 \frac{H_0 d_L}{c}=\frac{\alpha_0+\alpha_1 z
 +\cdots+\alpha_m z^m}{1+\beta_1 z+\cdots+\beta_n z^n}\,.
 \ee
 Note that the speed of light $c$ and Hubble constant $H_0$ are
 introduced from dimensional point of view. How to choose the
 order $(m,\,n)$ of the Pad\'e approximant is important. If the
 order is too low, the error of the Pad\'e approximant
 deviating from the real luminosity distance $d_L$ will be
 unacceptably large. If the order is too high, the number of
 free coefficients are too many and the uncertainties will be
 large. So, we choose a moderate order $(2,\,2)$ in this work,
 and then Eq.~(\ref{eq2}) becomes
 \be{eq3}
 D_L\equiv\frac{H_0 d_L}{c}=\frac{\alpha_0+\alpha_1 z+\alpha_2 z^2}
 {1+\beta_1 z+\beta_2 z^2}\,.
 \ee
 Obviously, it can work well in the whole redshift range
 $-1<z<\infty$, including not only the past but also the
 future of the universe. Especially, it is still finite even
 when $z\gg 1$. Also, it is easy to ensure the denominator
 not equal to zero at any redshift $z$ for suitable $\beta_1$
 and $\beta_2$. Thus, this parameterization based on the Pad\'e
 approximant can easily avoid the problems of the usual
 cosmography mentioned above. It is worth noting that this
 parameterization is a generalization of cosmography, since the
 Pad\'e approximant is a generalization of the Taylor series in
 fact.

\begin{center}
 \begin{figure}[t]
 \vspace{-15mm} 
 \centering
 \includegraphics[width=0.75\textwidth]{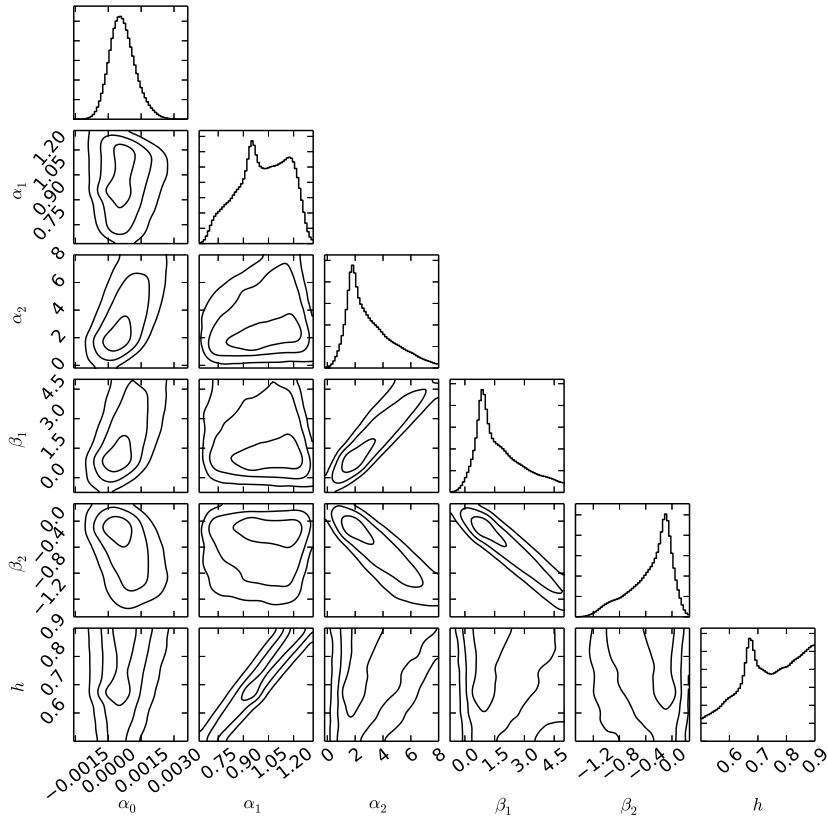}
 \caption{\label{fig1}
 The 1D marginalized distribution, and $1\sigma$, $2\sigma$,
 $3\sigma$ contours in the 2D model parameter spaces. These
 results are obtained by fitting the Pad\'e
 parameterization (\ref{eq3}) to SN data. See the text for details.}
 \end{figure}
 \end{center}

Naturally, it is important to confront the Pad\'e
 parameterization (\ref{eq3}) with the observational data,
 and see whether this parameterization works well. Since
 SNIa data are directly related to the luminosity distance, we
 can use them to constrain the parameterization (\ref{eq3}).
 Here, we consider the Union2.1 SNIa
 dataset~\cite{Suzuki:2011hu} consisting of 580 data points,
 which are given in terms of the distance modulus $\mu_{obs}(z_i)$.
 On the other hand, the theoretical distance modulus is given
 by~\cite{Weinberg2008,Riess:1998cb,1999ApJ...517..565P,
 Nesseris:2005ur,DiPietro:2002cz,Weidatafit}
 \be{eq4}
 \mu_{th}(z_i)=
 5\log_{10}\frac{d_L}{\rm Mpc}+25=5\log_{10}D_L(z_i)+\mu_0\,,
 \ee
 where $\mu_0\equiv 42.38-5\log_{10}h$, and $h$ is the Hubble
 constant $H_0$ in units of $100\,{\rm km/s/Mpc}$. In our case,
 $D_L$ has been given in Eq.~(\ref{eq3}). Correspondingly, the
 $\chi^2$ from 580 Union2.1 SNIa is given by
 \be{eq6}
 \chi^2_{SN}=\sum\limits_{i}\frac{\left[\,\mu_{obs}(z_i)-
 \mu_{th}(z_i)\,\right]^2}{\sigma^2(z_i)}\,
 \ee
 where $\sigma$ is the corresponding $1\sigma$ error. The
 best-fit model parameters are determined by minimizing
 $\chi^2$. In this work, we use the Markov chain Monte Carlo
 (MCMC) code {\it emcee}~\cite{ForemanMackey:2012ig} to find
 the best fits and the corresponding $68.3\%$, $95.4\%$ and
 $99.7\%$ confidence levels. We present the best-fit parameters
 with $1\sigma$, $2\sigma$, $3\sigma$ uncertainties and the
 corresponding $\chi^2_{min}$ in Table~\ref{tab1}. The 1D
 marginalized distribution, and $1\sigma$, $2\sigma$, $3\sigma$
 contours in the 2D model parameter spaces are also given in
 Fig.~\ref{fig1}.


 \begin{table}[t]
 \renewcommand{\arraystretch}{1.5}
 \begin{center}
 \vspace{-2mm} 
 \begin{tabular}{l|c|c} \hline\hline
   Dataset &  SN &  SN+CMB  \\ \hline
  $\chi^2_{min}$  & 562.530  &  562.171 \\ \hline
  $\chi^2_{min}/dof$~ & 0.980  & 0.978 \\ \hline
  $h$ & $0.67389_{-0.05318}^{+0.18215}\,(1\sigma)\,_{-0.14960}^{+0.22007}\,(2\sigma)\,_{-0.17192}^{+0.22578}\,(3\sigma)$ & \ \
   $0.69994_{-0.04318}^{+0.07525}\,(1\sigma)\,_{-0.08990}^{+0.09644}\,(2\sigma)\,_{-0.09942}^{+0.09987}\,(3\sigma)$ \ \\ \hline
  $\alpha_0$ & $0.00043_{-0.00033}^{+0.00080}\,(1\sigma)\,_{-0.00080}^{+0.00153}\,(2\sigma)\,_{-0.00127}^{+0.00226}\,(3\sigma)$ & \ \
  $0.00029_{-0.00033}^{+0.00055}\,(1\sigma)\,_{-0.00071}^{+0.00102}\,(2\sigma)\,_{-0.00109}^{+0.00153}\,(3\sigma)$ \ \\ \hline
  $\alpha_1$ & $0.94784_{-0.09201}^{+0.23285}\,(1\sigma)\,_{-0.22181}^{+0.30391}\,(2\sigma)\,_{-0.26775}^{+0.34522}\,(3\sigma)$ & \ \
   $0.98786_{-0.06788}^{+0.09886}\,(1\sigma)\,_{-0.13506}^{+0.14137}\,(2\sigma)\,_{-0.16705}^{+0.16747}\,(3\sigma)$ \ \\ \hline
   $\alpha_2$ & $1.74351_{-0.10466}^{+3.11237}\,(1\sigma)\,_{-0.87546}^{+5.24045}\,(2\sigma)_{-1.47114}^{+6.14252}\,(3\sigma)$ & \ \
   $1.42179_{-0.14728}^{+0.44075}\,(1\sigma)\,_{-0.42018}^{+0.86413}\,(2\sigma)\,_{-0.63735}^{+1.34014}\,(3\sigma)$ \ \\ \hline
   $\beta_1$ & $0.87449_{-0.14387}^{+2.28839}\,(1\sigma)_{-0.76494}^{+3.72218}\,(2\sigma)_{-1.24755}^{+4.09612}\,(3\sigma)$  & \ \
   $0.54097_{-0.10055}^{+0.24989}\,(1\sigma)\,_{-0.25922}^{+0.49820}\,(2\sigma)\,_{-0.39612}^{+0.78251}\,(3\sigma)$ \ \\ \hline
  $\beta_2$  & \ \ $-0.08709_{-0.57889}^{+0.04635}\,(1\sigma)_{-0.97493}^{+0.19247}\,(2\sigma)\,_{-1.18410}^{+0.29763}\,(3\sigma)$ \ \ & \ \
   $-0.00008_{-0.00011}^{+0.00007}\,(1\sigma)\,_{-0.00024}^{+0.00014}\,(2\sigma)\,_{-0.00038}^{+0.00020}\,(3\sigma)$ \ \\
 \hline\hline
 \end{tabular}
 \end{center}
 \caption{\label{tab1} The best-fit model parameters with
 $1\sigma$, $2\sigma$, $3\sigma$ uncertainties.
 The corresponding $\chi^2_{min}$ and $\chi^2_{min}/dof$ are
 also given. These results are obtained by fitting the
 Pad\'e parameterization (\ref{eq3}) to SN and SN+CMB data,
 respectively. See the text for details.}
 \end{table}

\begin{center}
 \begin{figure}[t]
 \vspace{-15mm} 
 \centering
 \includegraphics[width=0.75\textwidth]{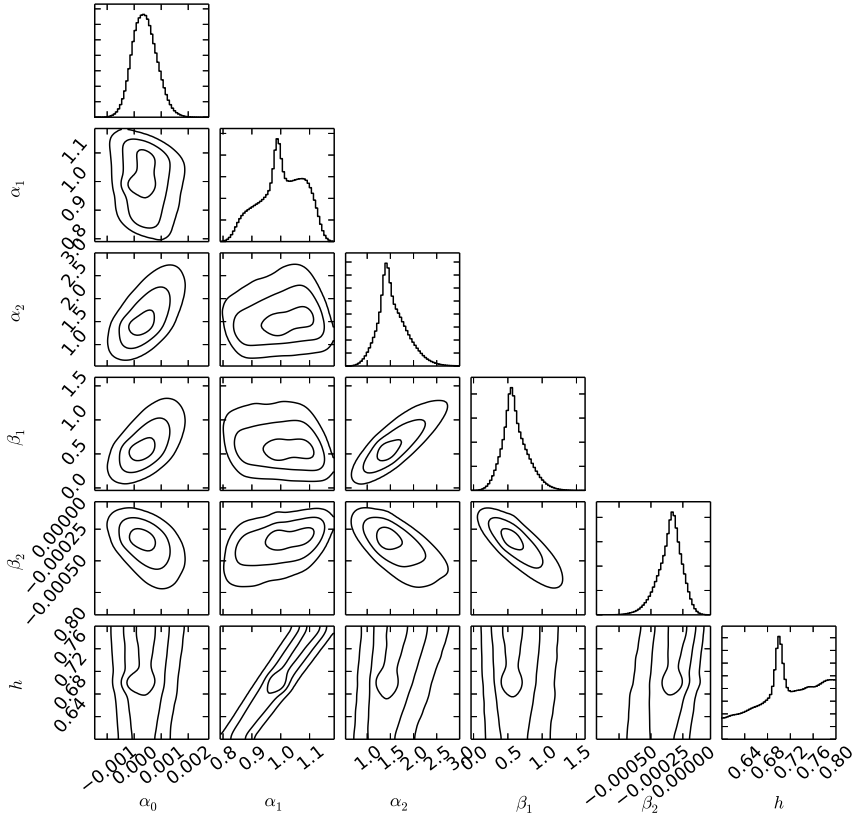}
 \caption{\label{fig2}
 The same as in Fig.~\ref{fig1}, except for SN+CMB data.}
 \end{figure}
 \end{center}

In addition to SNIa, the observation of cosmic microwave background
 (CMB) anisotropy~\cite{Ade:2015xua,Ade:2015rim} is another
 useful probe. However, using the full data of CMB  to perform
 a global fitting consumes a large amount of computation time
 and power. As an alternative, one can instead use the shift
 parameter $R$~\cite{Bond:1997wr} from CMB, which has been used
 extensively in the literature (including the works of the
 Planck and the WMAP Collaborations). It is
 argued in e.g.~\cite{Wang:2006ts,Wang:2013mha,Shafer:2013pxa} that
 the shift parameter $R$ is model-independent and contains the main
 information of the observation of CMB. As is well known, the
 shift parameter $R$ is defined
 by~\cite{Bond:1997wr,Ade:2015xua,Ade:2015rim,Wang:2006ts,
 Wang:2013mha,Shafer:2013pxa,Cai:2013owa}
 \be{eq9}
 R\equiv\sqrt{\Omega_{m0}H^2_0}\,\left(1+z_\ast\right)
 d_A(z_\ast)/c\,,
 \ee
 where $d_A(z)$ is the angular diameter distance to redshift
 $z$, which can be related to the luminosity distance $d_L$
 through (see e.g. the textbooks in~\cite{Weinberg2008})
 \be{eq10}
 d_A=\frac{d_L}{(1+z)^2}\,.
 \ee
 Noting $D_L\equiv H_0d_L/c$, we can recast Eq.~(\ref{eq9}) as
 \be{addeq4}
 R=\frac{\sqrt{\Omega_{m0}}\,D_L(z_\ast)}{1+z_\ast}\,,
 \ee
 and in our case, $D_L$ is given in Eq.~(\ref{eq3}). The redshift of
 the recombination $z_\ast=1089.90$, which was determined by
 the latest Planck 2015 data~\cite{Ade:2015xua}. $\Omega_{m0}$ is
 the present fractional density of pressureless matter, and it
 was determined as $\Omega_{m0}=0.308$ by the latest Planck
 2015 data~\cite{Ade:2015xua}. On the other hand, the observational
 value of $R$ has also been determined to be
 $R_{obs}=1.7382\pm 0.0088$ by the latest Planck
 2015 data~\cite{Ade:2015rim}. So, the $\chi^2$ from CMB is given by
 $\chi^2_{CMB}=(R-R_{obs})^2/\sigma_R^2$, and then the total
 $\chi^2$ from the combined SN+CMB data reads
 \be{eq11}
 \chi^2=\chi^2_{SN}+\chi^2_{CMB}\,,
 \ee
 where $\chi^2_{SN}$ is given in Eq.~(\ref{eq6}). Again, we use
 the MCMC code {\it emcee}~\cite{ForemanMackey:2012ig} to find
 the best fits and the corresponding $68.3\%$, $95.4\%$ and
 $99.7\%$ confidence levels to the combined SN+CMB data. We
 present the best-fit parameters with $1\sigma$, $2\sigma$,
 $3\sigma$ uncertainties and the corresponding $\chi^2_{min}$
 in the last column of Table~\ref{tab1}. The 1D marginalized
 distribution, and $1\sigma$, $2\sigma$, $3\sigma$ contours
 in the 2D model parameter spaces are also given in Fig.~\ref{fig2}.
 Thanks to CMB data, it is easy to see that the constraints on
 all parameters are significantly tightened (nb. Table~\ref{tab1}).
 By confronting the Pad\'e parameterization (\ref{eq3}) with
 the observational data, we see that this
 generalized cosmography works well.


\vspace{-17mm} 


\section{$y_\beta$-shift cosmography}\label{sec3}

In this section, we propose another generalization of
 cosmography inspired by the Pad\'e approximant. We stress that
 it is completely independent of the one proposed in the
 previous section. At first, we propose a new $y_\beta$-shift
 inspired by the Pad\'e approximant, and then derive the
 cosmography of the luminosity distance $d_L$ by expanding it
 as a Taylor series with respect to this new $y_\beta$-shift.
 We will also confront it with the observational data and see
 whether it works well.


\subsection{The formalism of $y_\beta$-shift cosmography}\label{sec3a}

At first, we give the motivation to propose the new $y_\beta$-shift.
 As is well known, the standard cosmography of the luminosity
 distance $d_L$ with respect to the redshift $z$
 is given by~\cite{Weinberg2008,Visser:2004bf,Bamba:2012cp,
 Dunsby:2015ers,Chiba:1998tc,Neben:2012wc}~(nb.~Eq.~(\ref{addeq3}))
 \be{eq12}
 \frac{H_0 d_L}{cz}=1+\frac{1}{2}\left(1-q_0\right)z+\dots\,.
 \ee
 One can instead parameterize it based on the
 Pad\'e approximant,
 \be{eq13}
 \frac{H_0 d_L}{cz}=\frac{\alpha_0+\alpha_1 z
 +\cdots+\alpha_m z^m}{1+\beta_1 z+\cdots+\beta_n z^n}\,.
 \ee
 Requiring that Eq.~(\ref{eq13}) coincides with Eq.~(\ref{eq12}) at
 $z\to 0$, we find that $\alpha_0\to 1$. So, we consider the Pad\'e
 approximant up to order $(1,\,1)$,
 \be{eq14}
 \frac{H_0d_L}{cz}=\frac{1+\alpha z}{1+\beta z}\,.
 \ee
 When $z\ll 1$, Eq.~(\ref{eq14}) can be expanded as
 \be{eq15}
 \frac{H_0 d_L}{cz}=
 1+(\alpha-\beta)z+{\cal O}\left(z^2\right)\,.
 \ee
 Comparing it with Eq.~(\ref{eq12}), we have
 $\alpha=(1-q_0)/2+\beta$. So, Eq.~(\ref{eq14}) becomes
 \be{eq17}
 \frac{H_0 d_L}{cz}=1+\frac{1}{2}\left(1-q_0
 \right)\frac{z}{1+\beta z}\,.
 \ee
 Putting Eqs.~(\ref{eq12}) and (\ref{eq17}) together, it is
 quite interesting to see that the quantity $z/(1+\beta z)$
 plays a role similar to the redshift $z$. This inspires us to
 propose a so-called $y_\beta$-shift,
 \be{addeq5}
 y_\beta\equiv\frac{z}{1+\beta z}\,,
 \ee
 where $\beta$ is a dimensionless constant. Obviously,
 $y_\beta=z$ if $\beta=0$, and $y_\beta=z/(1+z)$ if $\beta=1$.
 Noting that the redshift $z$~\cite{Weinberg2008,Visser:2004bf,
 Bamba:2012cp,Dunsby:2015ers,Chiba:1998tc,Neben:2012wc} and
 $y$-shift $y=z/(1+z)$~\cite{Cattoen:2007id,Cattoen:2008th} are
 extensively used~in~the~usual~cosmography, our $y_\beta$ can
 be regarded as their natural generalization. So, the
 cosmography with respect to $y_\beta$-shift is also a natural
 generalization of the usual cosmography in fact. We stress
 that the above discussions are only arguments to justify
 $y_\beta$, rather than strict derivations. Now, let us see how
 it might overcome the problems of the usual cosmography
 mentioned in Sec.~\ref{sec1}. First, $y_\beta$ is inspired by
 the Pad\'e approximant, and hence it is well motivated, not
 purely written by hand. Second, if $1-\beta\ll 1$, we
 have $y_\beta<1$ even for $z\gg 1$. So, the $y_\beta$-shift
 cosmography can converge safely. Third, it can describe the
 future evolution of the universe, if $1+\beta z\not=0$ in the
 redshift range $-1<z<0$. In fact, to ensure $y_\beta$ remains
 regular in not only the past but also in the future of the
 universe ($-1<z<\infty$), it is required that
 \be{addeq6}
 0<\beta<1\,.
 \ee

In the following, let us derive the cosmography of the
 luminosity distance $d_L$ with respect to $y_\beta$-shift.
 As in e.g.~\cite{Weinberg2008,Visser:2004bf,Bamba:2012cp,
 Dunsby:2015ers,Chiba:1998tc,Neben:2012wc,Cattoen:2007id,
 Cattoen:2008th}, it is convenient to introduce the following
 functions,
 \bea
 H(t) &\equiv & +\frac{1}{a}\frac{da}{dt}\,,\label{eq18}\\
 q(t) &\equiv & -\frac{1}{aH^2}\frac{d^{2}a}{dt^{2}}\,,\label{eq19}\\
 j(t) &\equiv & +\frac{1}{aH^3}\frac{d^{3}a}{dt^{3}}\,,\label{eq20}\\
 s(t) &\equiv & +\frac{1}{aH^4}\frac{d^{4}a}{dt^{4}}\,,\label{eq21}\\
 l(t) &\equiv & +\frac{1}{aH^5}\frac{d^{5}a}{dt^{5}}\,,\label{eq22}
 \eea
 which are usually referred to as the Hubble, deceleration,
 jerk, snap, and lerk parameters. Using these definitions, the
 Taylor series expansion of scale factor $a$ up to 5th order
 around the time $t_0$ reads
\be{eq23}
\begin{split}
a(t)&=a(t_{0})\left[1 + H_{0} (t-t_{0}) - \frac{q_{0}}{2}
H_{0}^{2} (t-t_{0})^{2} + \frac{j_{0}}{3!} H_{0}^{3}(t-t_{0})^{3}\right.\\
&\left.\quad + \frac{s_{0}}{4!} H_{0}^{4} (t-t_{0})^{4}
+\frac{l_{0}}{5!} H_{0}^{5} (t-t_{0})^{5}
+\mathcal{O}\left((t-t_{0})^{6}\right)\right]\,,
\end{split}
\ee
 where the subscript ``0'' indicates the value of the corresponding
 quantity evaluated at the time $t_0$. The physical distance
 traveled by a photon that is emitted at the time $t$ and
 absorbed at the current epoch $t_0$ is given by
 \be{eq24}
 D = c \int d\tilde{t}=c\left(t_0 - t\right)\,,
 \ee
 where the time difference $\delta t=t_0-t$ is called the
 ``lookback time''. So, we have
 \be{eq25}
 1 + z = \frac{a(t_0)}{a(t)}=
 \frac{a(t_0)}{a(t_0-\delta t)}=\frac{a(t_0)}{a(t_0-D/c)}\,.
 \ee
 Using Eq.~(\ref{eq23}), its right-hand side can
 be expanded as a Taylor series with respect to $H_0 D/c\,$,
\begin{align}
\frac{a(t_0)}{a(t_0-D/c)}&=1 + \frac{H_0 D}{c} + \left(1+
\frac{1}{2}q_0\right)\left(\frac{H_0 D}{c}\right)^2+\left(1+
q_0+\frac{1}{6}j_0\right)\left(\frac{H_0 D}{c}\right)^3 \nonumber\\
&+\left(1+\frac{3}{2}q_0+\frac{1}{4}q_0^2+\frac{1}{3}j_0-
\frac{1}{24}s_0\right)\left(\frac{H_0 D}{c}\right)^4\nonumber\\
&+\left(1+2q_0+\frac{3}{4}q_0^2+\frac{1}{6}q_0j_0
+\frac{1}{2}j_0-\frac{1}{12}s_0+\frac{1}{120}l_0\right)
\left(\frac{H_0 D}{c}\right)^5+\mathcal{O}\left(\left(
\frac{H_0 D}{c}\right)^6\right)\,.\label{eq26}
\end{align}
Therefore, from Eqs.~(\ref{eq25}) and (\ref{eq26}), we obtain
\be{eq27}
\begin{split}
z(D)=\mathcal{Z}_D^{(1)}\left(\frac{H_0 D}{c}\right) & +
\mathcal{Z}_D^{(2)}\left(\frac{H_0 D}{c}\right)^2
+\mathcal{Z}_D^{(3)}\left(\frac{H_0 D}{c}\right)^3 \\
&+\mathcal{Z}_D^{(4)}\left(\frac{H_0 D}{c}\right)^4
+\mathcal{Z}_D^{(5)}\left(\frac{H_0 D}{c}\right)^5
+\mathcal{O}\left(\left(\frac{H_0 D}{c}\right)^6\right)\,,
\end{split}
\ee
where
\bea
&&\mathcal{Z}_D^{(1)}=1\,, \label{eq28}\\
&&\mathcal{Z}_D^{(2)}=1+\frac{1}{2}q_0\,, \label{eq29}\\
&&\mathcal{Z}_D^{(3)}=1+q_0+\frac{1}{6}j_0\,, \label{eq30}\\
&&\mathcal{Z}_D^{(4)}=1+\frac{3}{2}q_0+\frac{1}{4}q_0^2+
\frac{1}{3}j_0-\frac{1}{24}s_0\,, \label{eq31}\\
&&\mathcal{Z}_D^{(5)}=1+2q_0+\frac{3}{4}q_0^2
+\frac{1}{6}q_0j_0+\frac{1}{2}j_0-\frac{1}{12}s_0
+\frac{1}{120}l_0 \,. \label{eq32}
\eea
We can expand $D$ as a Taylor series up to 5th order with
 respect to $y_\beta$,
\be{eq33}
D(y_\beta)=\frac{c}{H_0}\left[\mathcal{D}_y^{(1)}y_\beta+
\mathcal{D}_y^{(2)}y_\beta^2+\mathcal{D}_y^{(3)}y_\beta^3+
\mathcal{D}_y^{(4)}y_\beta^4+\mathcal{D}_y^{(5)}y_\beta^5+
\mathcal{O}\left(y_\beta^6\right)\right]\,,
\ee
in which we have used $D(y_\beta=0)=D(t=t_0)=0$
 from Eq.~(\ref{eq24}). Substituting Eq.~(\ref{eq33}) into
 Eq.~(\ref{eq27}), and using $z=y_\beta/(1-\beta y_\beta)$
 obtained from Eq.~(\ref{addeq5}), we have
\be{eq34}
\begin{split}
\frac{y_\beta}{1-\beta y_\beta}
&=\mathcal{Z}_D^{(1)}\left(\mathcal{D}_y^{(1)}y_\beta+
\mathcal{D}_y^{(2)}y_\beta^2+\mathcal{D}_y^{(3)}y_\beta^3+
\mathcal{D}_y^{(4)}y_\beta^4+\mathcal{D}_y^{(5)}y_\beta^5+
\mathcal{O}\left(y_\beta^6\right)\right)\\
&\quad+\mathcal{Z}_D^{(2)}\left(\mathcal{D}_y^{(1)}y_\beta+
\mathcal{D}_y^{(2)}y_\beta^2+\mathcal{D}_y^{(3)}y_\beta^3+
\mathcal{D}_y^{(4)}y_\beta^4+\mathcal{D}_y^{(5)}y_\beta^5+
\mathcal{O}\left(y_\beta^6\right)\right)^2\\
&\quad+\mathcal{Z}_D^{(3)}\left(\mathcal{D}_y^{(1)}y_\beta+
\mathcal{D}_y^{(2)}y_\beta^2+\mathcal{D}_y^{(3)}y_\beta^3+
\mathcal{D}_y^{(4)}y_\beta^4+\mathcal{D}_y^{(5)}y_\beta^5+
\mathcal{O}\left(y_\beta^6\right)\right)^3\\
&\quad+\mathcal{Z}_D^{(4)}\left(\mathcal{D}_y^{(1)}y_\beta+
\mathcal{D}_y^{(2)}y_\beta^2+\mathcal{D}_y^{(3)}y_\beta^3+
\mathcal{D}_y^{(4)}y_\beta^4+\mathcal{D}_y^{(5)}y_\beta^5+
\mathcal{O}\left(y_\beta^6\right)\right)^4\\
&\quad+\mathcal{Z}_D^{(5)}\left(\mathcal{D}_y^{(1)}y_\beta+
\mathcal{D}_y^{(5)}y_\beta^2+\mathcal{D}_y^{(3)}y_\beta^3+
\mathcal{D}_y^{(4)}y_\beta^4+\mathcal{D}_y^{(5)}y_\beta^5+
\mathcal{O}\left(y_\beta^6\right)\right)^5\,.\\
\end{split}
\ee
After some algebra, Eq.~(\ref{eq34}) becomes
\begin{align}
0&=\left(\mathcal{D}_y^{(1)}-1\right)y_\beta+
\left[\mathcal{D}_y^{(2)}-\beta\mathcal{D}_y^{(1)}+\left(
\mathcal{D}_y^{(1)}\right)^2 \mathcal{Z}_D^{(2)}\right]y_\beta^2\nonumber\\
&\quad+\left\{\mathcal{D}_y^{(3)}-\beta\mathcal{D}_y^{(2)}
+\mathcal{Z}_D^{(2)}\left[2\mathcal{D}_y^{(1)}\mathcal{D}_y^{(2)}-
\beta\left(\mathcal{D}_y^{(1)}\right)^2\right]
+\mathcal{Z}_D^{(3)}\left(\mathcal{D}_y^{(1)}\right)^3
\right\}y_\beta^3\nonumber\\
&\quad+\left\{\mathcal{D}_y^{(4)}-\beta\mathcal{D}_y^{(3)}+
\mathcal{Z}_D^{(2)}\left[\left(\mathcal{D}_y^{(2)}\right)^2+
2\mathcal{D}_y^{(1)}\mathcal{D}_y^{(3)}-
2\beta\mathcal{D}_y^{(1)}\mathcal{D}_y^{(2)}\right]\right.\nonumber\\
&\quad\quad\left.+\mathcal{Z}_D^{(3)}\left[3\left(
\mathcal{D}_y^{(1)}\right)^2\mathcal{D}_y^{(2)}-
\beta\left(\mathcal{D}_y^{(1)}\right)^3\right]+\mathcal{Z}_D^{(4)}
\left(\mathcal{D}_y^{(1)}\right)^4\right\}y_\beta^4\nonumber\\
&\quad+\left\{\mathcal{D}_y^{(5)}-\beta\mathcal{D}_y^{(4)}+
\mathcal{Z}_D^{(2)}\left[2\left(\mathcal{D}_y^{(2)}\mathcal{D}_y^{(3)}
+\mathcal{D}_y^{(1)}\mathcal{D}_y^{(4)}\right)-
\beta\left(\left(\mathcal{D}_y^{(2)}\right)^2
+2\mathcal{D}_y^{(1)}\mathcal{D}_y^{(3)}\right)\right]\right.\nonumber\\
&\quad\quad+3\mathcal{Z}_D^{(3)}\left[\mathcal{D}_y^{(1)}
\left(\mathcal{D}_y^{(2)}\right)^2+\left(\mathcal{D}_y^{(1)}
\right)^2\mathcal{D}_y^{(3)}-\beta\left(\mathcal{D}_y^{(1)}
\right)^2\mathcal{D}_y^{(2)}\right]\nonumber\\
&\quad\quad\left.+\mathcal{Z}_D^{(4)}\left(\mathcal{D}_y^{(1)}
\right)^3\left(4\mathcal{D}_y^{(2)}-\beta\mathcal{D}_y^{(1)}\right)
+\mathcal{Z}_D^{(5)}\left(\mathcal{D}_y^{(1)}\right)^5
\right\}y_\beta^5+\mathcal{O}\left(y_\beta^6\right)\,,\label{eq35}
\end{align}
in which we have used ${\cal Z}_D^{(1)}=1$ from
 Eq.~(\ref{eq28}). Requiring all the coefficients
 of $y_\beta^i$ in Eq.~(\ref{eq35}) to be zero, and using
 Eqs.~(\ref{eq29})---(\ref{eq32}), we find that
\begin{align}
\mathcal{D}_y^{(1)}&=1\,, \label{eq36}\\
\mathcal{D}_y^{(2)}&=-\left(1+\frac{1}{2}q_0-\beta\right)\,,
 \label{eq37}\\
\mathcal{D}_y^{(3)}&=(\beta-1)^2-\left(\beta-1\right)q_0+
\frac{1}{2}q_0^2-\frac{1}{6}j_0\,, \label{eq38}\\[1mm]
\mathcal{D}_y^{(4)}&=-1+3\beta-3\beta^2+\beta^3-\left(
\frac{3}{2}\beta^2-3\beta+\frac{3}{2}\right)q_0-
\frac{3}{2}\left(1-\beta\right)q_0^2\notag \\
&\quad-\frac{5}{8}q_0^3+\frac{5}{12}q_0j_0-\frac{1}{2}\left(
\beta-1\right)j_0+\frac{1}{24}s_0\,, \label{eq39}\\[1mm]
\mathcal{D}_y^{(5)}&=1-4\beta+6\beta^2-4\beta^3+\beta^4+
2\left(1-3\beta+3\beta^2-\beta^3\right)q_0+
3\left(1-2\beta+\beta^2\right)q_0^2\notag\\
&\quad+\frac{5}{2}\left(1-\beta\right)q_0^3+\frac{7}{8}q_0^2j_0
+\frac{5}{3}\left(\beta-1\right)q_0j_0-(\beta-1)^2j_0\notag\\
&\quad+\frac{1}{12}j_0^2+\frac{1}{6}\left(\beta-1\right)s_0-
\frac{1}{8}q_0s_0-\frac{1}{120}l_0\,.\label{eq40}
\end{align}
The role of the observable physical quantity is played by
 the luminosity distance. Let the photon be emitted at
 $r$-coordinate $r=0$ at the time $t$, and absorbed at
 $r$-coordinate $r=r_0$ at the time $t_0$. Then,
 the luminosity distance $d_L$ is given by (see e.g. the
 textbooks in~\cite{Weinberg2008})
 \be{eq41}
 d_L = \frac{a(t_0)}{a(t_0-D/c)}\,\left(a(t_0)\,r_0\right)\,.
 \ee
To calculate $d_L(D)$, we need $r_0(D)$. From
 Eq.~(\ref{addeq1}), it is easy to
 obtain~\cite{Visser:2004bf,Weinberg2008}
 \be{eq42}
 r_0(D) = \begin{cases}
  \disp\sin \left( \int_{t_0- D/c}^{t_0}\, \frac{c\,dt}{a(t)}
  \right) & \ {\rm for}\ \ k = +1\,, \\
  &  \\
  \disp\int_{t_0- D/c}^{t_0}\, \frac{c\,dt}{a(t)}
  & \ {\rm for}\ \ k = 0\,, \\
  &  \\
  \disp\sinh \left( \int_{t_0- D/c}^{t_0}\, \frac{c\,dt}{a(t)}
  \right) & \ {\rm for}\ \ k = -1\,.
 \end{cases}
 \ee
Using Eq.~(\ref{eq23}), we can calculate the
 integration~\cite{Bamba:2012cp}
\begin{align}
\int_{t_0- D/c}^{t_0}\,\frac{c\,dt}{a(t)} &=\frac{D}{a_0}
\left\{1+\frac{1}{2}\left(\frac{H_0 D}{c}\right)+
\frac{1}{6}\left(2+q_0\right)\left(\frac{H_0 D}{c}\right)^2
+\frac{1}{24}\Big[6\big(1+q_0\big)+
j_0\Big]\left(\frac{H_0D}{c}\right)^3\right.\nonumber\\
&\quad+\left.\frac{1}{120}\left(24+36q_0+6q_0^2+
8j_0-s_0\right)\left(\frac{H_0 D}{c}\right)^4+\mathcal{O}\left(
\left(\frac{H_0 D}{c}\right)^5\right)\right\}\,,\label{eq44}
\end{align}
where $a_0=a(t_0)$. At first glance, we should deal with three
 cases with space curvature $k=+1$, $0$, $-1$ separately.
 Fortunately, we need not to do so. As is well known, the
 Taylor series expansions of $\sin x$ and $\sinh x$ are given
 by $\sin x=x-x^3/3!+x^5/5!-x^7/7!+\dots$ and
 $\sinh x=x+x^3/3!+x^5/5!+x^7/7!+\dots$, respectively. Noting
 that $r_0(D)\to\sin(\ast)$ for $k=+1$, and
 $r_0(D)\to\sinh(\ast)$ for $k=-1$, we can write
 the Taylor series expansion of $r_0(D)$ in
 Eq.~(\ref{eq42}) as a uniform expression for $k=+1$, $0$,
 $-1$, i.e.~\cite{Bamba:2012cp,Visser:2004bf}
 \be{addeq7}
 r_0(D)=\left(\int_{t_0- D/c}^{t_0}\,\frac{c\,dt}{a(t)}
 \right)-\frac{k}{3!}\left(\int_{t_0-D/c}^{t_0}
 \,\frac{c\,dt}{a(t)}\right)^3+\frac{k^2}{5!}\left(\int_{t_0
 -D/c}^{t_0}\,\frac{c\,dt}{a(t)}\right)^5
 +{\cal O}\left(\left(\int_{t_0-D/c}^{t_0}\,\frac{c\,dt}{a(t)}
 \right)^7\right).
 \ee
Substituting Eq.~(\ref{eq44}) into Eq.~(\ref{addeq7}), we have
\begin{align}
r_0(D)&=\frac{D}{a_0}\left\{1+\frac{1}{2}\left(\frac{H_0 D}{c}\right)
+\frac{1}{6}\left(2+q_0-\frac{kc^2}{H_0^2a_0^2}\right)\left(
\frac{H_0 D}{c}\right)^2\right.
+\frac{1}{24}\left[6\left(1+q_0\right)+j_0-\frac{6kc^2}{H_0^2 a_0^2}
\right]\left(\frac{H_0 D}{c}\right)^3\nonumber\\
&\quad +\frac{1}{120}\left[24+36q_0+6q_0^2+8j_0-
s_0-\frac{5kc^2\left(7+2q_0\right)}{H_0^2 a_0^2}
+\left(\frac{kc^2}{H_0^2 a_0^2}\right)^2\right]
\left(\frac{H_0 D}{c}\right)^4\nonumber\\
&\quad +\left.\mathcal{O}\left(\left(\frac{H_0D}{c}\right)^5
\right)\right\}\,.\label{addeq8}
\end{align}
Substituting Eqs.~(\ref{eq26}), (\ref{addeq8}), (\ref{eq33})
 with (\ref{eq36})---(\ref{eq40}) into Eq.~(\ref{eq41}), we
 finally obtain the cosmography of the luminosity
 distance $d_L$ with respect to $y_\beta$-shift,
\be{eq45}
d_L(y_\beta)=\frac{c}{H_0}\Big[\mathcal{D}_L^{(1)}\,y_\beta+
\mathcal{D}_L^{(2)}\,y_\beta^2+\mathcal{D}_L^{(3)}\,y_\beta^3
+\mathcal{D}_L^{(4)}\,y_\beta^4+\mathcal{D}_L^{(5)}\,y_\beta^5
+\mathcal{O}\left(y_\beta^6\right)\Big]\,,
\ee
where
\begin{align}
\mathcal{D}_L^{(1)} & =1\,,\label{eq46}\\
\mathcal{D}_L^{(2)} & =\frac{1}{2}\left(1-q_0\right)+
\beta\,,\label{eq47}\\
\mathcal{D}_L^{(3)} & =-\frac{1}{6}+\beta+\beta^2+\left(\frac{1}{6}
-\beta\right)q_0+\frac{1}{2}q_0^2-\frac{1}{6}j_0-\frac{1}{6}
\left(\frac{kc^2}{H_0^2 a_0^2}\right)\,,\label{eq48}\\
\mathcal{D}_L^{(4)} & =\frac{1}{12}
-\frac{1}{2}\beta+\frac{3}{2}\beta^2+\beta^3+\left(\frac{1}{2}\beta
-\frac{3}{2}\beta^2-\frac{1}{12}\right)q_0+\left(\frac{3}{2}\beta-
\frac{5}{8}\right)q_0^2-\frac{5}{8}q_0^3\notag\\
&\quad+\frac{5}{12}q_0j_0+\left(\frac{5}{24}-\frac{1}{2}\beta
\right)j_0+\frac{1}{24}s_0+\frac{1}{12}\left(1+3q_0-6\beta
\right)\left(\frac{kc^2}{H_0^2 a_0^2}\right)\,,\label{eq49}\\
\mathcal{D}_L^{(5)} & =-\frac{1}{20}+\frac{1}{3}\beta-\beta^2
+2\beta^3+\beta^4+\left(\frac{1}{20}-\frac{1}{3}\beta+
\beta^2-2\beta^3\right)q_0\notag\\
&\quad+\left(\frac{27}{40}-\frac{5}{2}\beta
+3\beta^2\right)q_0^2+\left(\frac{11}{8}-
\frac{5}{2}\beta\right)q_0^3+\frac{7}{8}q_0^4+
\left(\frac{5}{3}\beta-\frac{11}{12}\right)q_0j_0\notag\\
&\quad-\frac{7}{8}q_0^2j_0+\left(\frac{5}{6}\beta-\beta^2
-\frac{9}{40}\right)j_0+\frac{1}{12}j_0^2+
\left(\frac{1}{6}\beta-\frac{11}{120}\right)s_0-
\frac{1}{8}q_0s_0-\frac{1}{120}l_0\notag\\
&\quad-\frac{1}{24}\left(1+8q_0+9q_0^2-2j_0
-8\beta-24\beta q_0+24\beta^2\right)\left(
\frac{kc^2}{H_0^2 a_0^2}\right)
+\frac{1}{120}\left(\frac{kc^2}{H_0^2 a_0^2}\right)^2\,.\label{eq50}
\end{align}
One can check that if $\beta=0$ or $\beta=1$, these results
 reduce to the one of cosmography with respect to redshift
 $z$ (e.g.~\cite{Bamba:2012cp,Visser:2004bf}) or $y$-shift
 $y=z/(1+z)$ (e.g.~\cite{Cattoen:2007id,Cattoen:2008th,
 Bamba:2012cp,Vitagliano:2009et}), respectively.
 The cosmography with respect to $y_\beta$-shift
 $y_\beta=z/(1+\beta z)$ obtained here is more general.


\subsection{Cosmological constraints from the observational data}\label{sec3b}

It is natural to confront our new cosmography with the
 observational data, and see whether this new

\newpage 


 \begin{center}
 \begin{figure}[t]
 \vspace{-15mm} 
 \centering
 \includegraphics[width=0.75\textwidth]{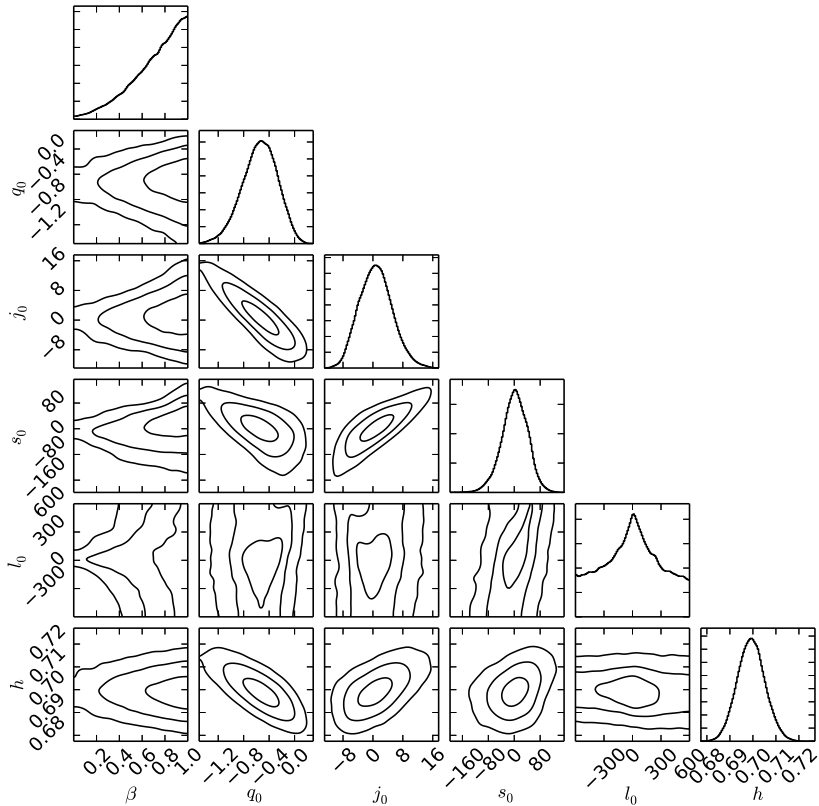}
 \caption{\label{fig3}
 The 1D marginalized distribution, and $1\sigma$, $2\sigma$,
 $3\sigma$ contours in the 2D model parameter spaces. These
 results are obtained by fitting the
 $y_\beta$-shift cosmography (\ref{eq45}) to SN data with
 the prior $0<\beta<1$. See the text for details.}
 \end{figure}
 \end{center}

\begin{center}
 \begin{figure}[tb]
 \vspace{-12.99mm} 
 \centering
 \includegraphics[width=0.75\textwidth]{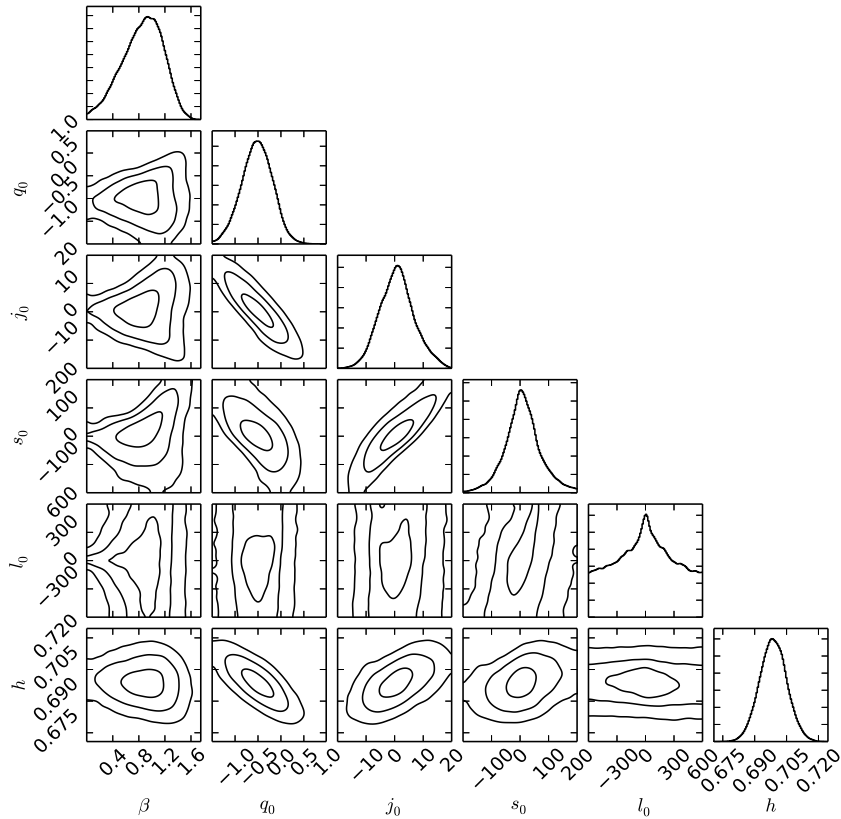}
 \caption{\label{fig4}
 The same as in Fig.~\ref{fig3}, except for the prior $0<\beta<2$.}
 \end{figure}
 \end{center}

\vspace{-8mm} 

\noindent cosmography
 works well. Here, we only consider a flat FRW universe with
 \be{addeq9}
 k=0\,.
 \ee
 Similar to Sec.~\ref{sec2}, we first
 consider the constraints from the Union2.1 SNIa
 dataset~\cite{Suzuki:2011hu}. The $\chi^2$ from 580 Union2.1
 SNIa is given in Eq.~(\ref{eq6}), in which $\mu_{th}$ is given
 by Eq.~(\ref{eq4}), $D_L\equiv H_0 d_L/c$ and $d_L$ is given
 by Eq.~(\ref{eq45}) in our case. Again, we use the MCMC code
 {\it emcee}~\cite{ForemanMackey:2012ig} to find the best fits
 and the corresponding $68.3\%$, $95.4\%$ and $99.7\%$
 confidence levels to SN data. Note that the prior $0<\beta<1$
 is required in Eq.~(\ref{addeq6}). We present the best-fit
 parameters with $1\sigma$, $2\sigma$, $3\sigma$ uncertainties
 and the corresponding $\chi^2_{min}$ in Table~\ref{tab2}. The
 1D marginalized distribution, and $1\sigma$, $2\sigma$,
 $3\sigma$ contours in the 2D model parameter spaces are also
 given in Fig.~\ref{fig3}. It is easy to see that the parameter
 $\beta$ cannot be well constrained, and its 1D marginalized
 distribution is not Gaussian. So, we temporarily relax the
 prior to $0<\beta<2$. In this case, the best-fit parameters
 with $1\sigma$, $2\sigma$, $3\sigma$ uncertainties and the
 corresponding $\chi^2_{min}$ are given in the last column of
 Table~\ref{tab2}, while the 1D marginalized distribution,
 and $1\sigma$, $2\sigma$, $3\sigma$ contours in the 2D model
 parameter spaces are also given in Fig.~\ref{fig4}. Now, the
 constraint on $\beta$ looks well. The following discussions
 hold for both cases with $0<\beta<1$ and $0<\beta<2$ in fact.
 It is easy to see that SN data favor a best-fit $\beta$ close
 to $1$, and hence $y_\beta$ is close to the usual $y$-shift
 $y=z/(1+z)$. However, $\beta$ can still significantly deviate
 from $1$ in the wide $1\sigma$, $2\sigma$, $3\sigma$ regions.
 In particular, $\beta$ can even be close to $0$ in the
 $3\sigma$ region. On the other hand, we find that $q_0<0$
 at $1\sigma$ confidence level, which indicates the cosmic
 expansion is accelerating (nb. the definition in
 Eq.~(\ref{eq19})). Of course, it is not surprising that the
 constraints on $l_0$ and $s_0$ are fairly loose, since we
 use only SN data here.

\begin{center}
 \begin{figure}[tb]
 \vspace{-12.99mm} 
 \centering
 \includegraphics[width=0.75\textwidth]{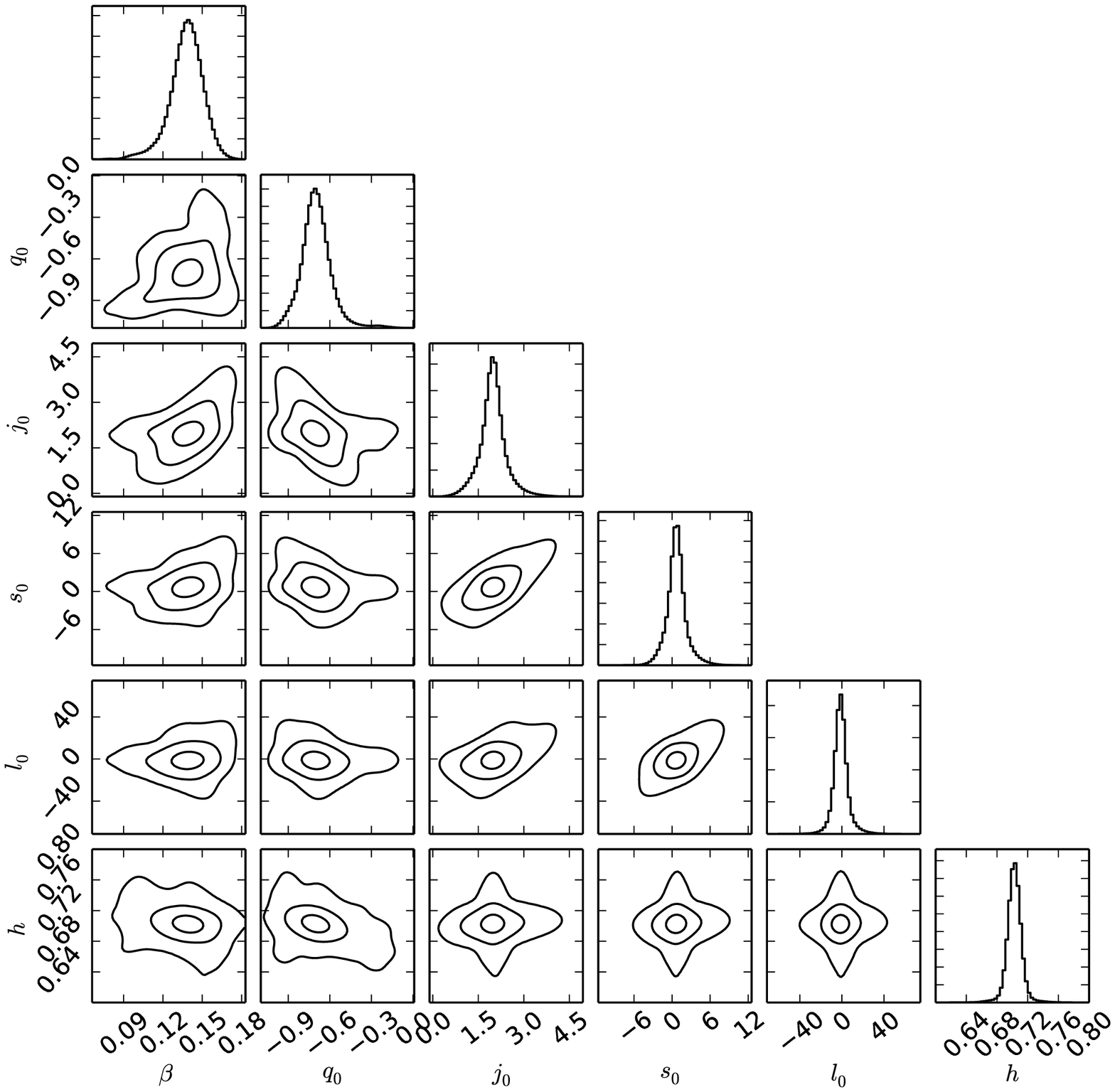}
 \caption{\label{fig5}
 The same as in Fig.~\ref{fig3}, except for SN+CMB data, and
 the prior $0<\beta<1$ is required.}
 \end{figure}
 \end{center}

Let us further consider the constraints from the combined
 SN+CMB data. The corresponding $\chi^2$ is given
 by Eq.~(\ref{eq11}), in which $\chi^2_{SN}$ is given by
 Eq.~(\ref{eq6}) and $\chi^2_{CMB}=(R-R_{obs})^2/\sigma_R^2$,
 while $D_L\equiv H_0 d_L/c$ and $d_L$ is given by Eq.~(\ref{eq45})
 in our case. Note that the prior $0<\beta<1$ is still required
 in Eq.~(\ref{addeq6}). We present the best-fit parameters with
 $1\sigma$, $2\sigma$, $3\sigma$ uncertainties and the
 corresponding $\chi^2_{min}$ in Table~\ref{tab3}. The 1D
 marginalized distribution, and $1\sigma$, $2\sigma$, $3\sigma$
 contours in the 2D model parameter spaces are also given in
 Fig.~\ref{fig5}. Obviously, the constraints on all parameters
 are significantly tightened, mainly thanks to the CMB data. It
 is interesting to see that the combined SN+CMB data favor
 a best-fit $\beta$ close to $0$, and hence $y_\beta$ is close
 to the usual redshift $z$. Noting that SN data favors a
 best-fit $\beta$ close to $1$ as mentioned above, this
 indicates that there is tension between the SN and the CMB
 data. However, we stress that the constraints from SN and
 SN+CMB data are still consistent within the $3\sigma$
 confidence level. On the other hand, we see that $q_0<0$,
 $j_0>0$ beyond $3\sigma$ confidence level, which strongly
 indicates the cosmic expansion is accelerating, and the
 acceleration is increasing (nb.
 the definitions in Eqs.~(\ref{eq19}) and (\ref{eq20})).


 \begin{table}[tb]
 \renewcommand{\arraystretch}{1.5}
 \begin{center}
 \vspace{-2mm} 
 \begin{tabular}{l|c|c} \hline\hline
   Dataset &  SN ($0< \beta <1$) &  SN ($0< \beta <2$)  \\ \hline
   $\chi^2_{min}$ & 565.317 &  564.668  \\ \hline
  $\chi^2_{min}/dof$~ &  0.985  &  0.984 \ \\ \hline
  $\beta$ &  $<1$ &
  $0.94516_{-0.42111}^{+0.21990}\,(1\sigma)\,_{-0.75642}^{+0.44858}\,(2\sigma)\,_{-0.92425}^{+0.63060}\,(3\sigma)$  \\ \hline
  $h$ & $0.69932_{-0.00600}^{+0.00589}\,(1\sigma)\,_{-0.01170}^{+0.01221}\,(2\sigma)\,_{-0.01709}^{+0.01819}\,(3\sigma)$ &
   $0.69818_{-0.00581}^{+0.00682}\,(1\sigma)\,_{-0.01208}^{+0.01338}\,(2\sigma)\,_{-0.01847}^{+0.01938}\,(3\sigma)$  \\ \hline
  $q_0$ & \ $-0.55618_{-0.26980}^{+0.28353}\,(1\sigma)\,_{-0.57862}^{+0.50908}\,(2\sigma)_{-0.85393}^{+0.68637}\,(3\sigma)$ \ &
   $-0.52943_{-0.32969}^{+0.34475}\,(1\sigma)\,_{-0.67578}^{+0.65817}\,(2\sigma)\,_{-0.91769}^{+1.01555}\,(3\sigma)$  \\ \hline
  $j_0$ &  $0.55909_{-4.21374}^{+4.59544}\,(1\sigma)_{-7.76870}^{+9.48256}\,(2\sigma)_{-11.10038}^{+14.08874}\,(3\sigma)$  &
   \,\ $0.39037_{-5.56827}^{+6.47105}\,(1\sigma)\,_{-11.19167}^{+13.46629}\,(2\sigma)\,_{-16.72432}^{+17.82005}\,(3\sigma)$ \,\ \\ \hline
  $s_0$ & $1.90756_{-39.1773}^{+38.9127}\,(1\sigma)_{-87.1962}^{+75.4962}\,(2\sigma)_{-151.8190}^{+116.3544}\,(3\sigma)$ &
   $3.97831_{-52.4932}^{+63.2142}\,(1\sigma)\,_{-127.5164}^{+145.1346}\,(2\sigma)\,_{-191.6690}^{+191.0794}\,(3\sigma)$  \\ \hline
  $l_0$ & $-6.32885_{-342.262}^{+344.384}\,(1\sigma)_{-554.510}^{+563.219}\,(2\sigma)\,_{-591.835}^{+604.266}\,(3\sigma)$  &
   $0.02741_{-360.915}^{+357.489}\,(1\sigma)\,_{-562.618}^{+562.259}\,(2\sigma)\,_{-597.821}^{+597.730}\,(3\sigma)$  \\
 \hline\hline
 \end{tabular}
 \end{center}
 \caption{\label{tab2} The best-fit model parameters with
 $1\sigma$, $2\sigma$, $3\sigma$ uncertainties.
 The corresponding $\chi^2_{min}$ and $\chi^2_{min}/dof$ are
 also given. These results are obtained by fitting the
 $y_\beta$-shift cosmography (\ref{eq45}) to SN data with the
 priors $0<\beta<1$ and $0<\beta<2$, respectively. See the
 text for details.}
 \end{table}



\begin{table}[tb]
 \renewcommand{\arraystretch}{1.5}
 \begin{center}
 \vspace{4mm} 
 \begin{tabular}{l|cc} \hline\hline
  Dataset &  SN+CMB  \\ \hline
  $\chi^2_{min}$ &  562.654  \\ \hline
  $\chi^2_{min}/dof$ \ \ &  0.979  \\ \hline
  $\beta$ & \ \ $0.13891_{-0.01202}^{+0.01050}\,(1\sigma)\,_{-0.03229}^{+0.02126}\,(2\sigma)\,_{-0.05907}^{+0.03294}\,(3\sigma)$ \\ \hline
  $h$ & \ \ $0.70236_{-0.00556}^{+0.00611}\,(1\sigma)\,_{-0.01938}^{+0.01844}\,(2\sigma)\,_{-0.07461}^{+0.07016}\,(3\sigma)$ \\ \hline
  $q_0$ & \ \ $-0.70925_{-0.07971}^{+0.09912}\,(1\sigma)\,_{-0.19356}^{+0.27603}\,(2\sigma)_{-0.27210}^{+0.55496}\,(3\sigma)$  \\ \hline
  $j_0$ & \ \ $1.98747_{-0.35293}^{+0.31463}\,(1\sigma)_{-0.92251}^{+1.03983}\,(2\sigma)_{-1.50048}^{+2.07407}\,(3\sigma)$  \\ \hline
  $s_0$ & \ \ $0.69908_{-1.11125}^{+0.99476}\,(1\sigma)_{-3.04250}^{+3.71175}\,(2\sigma)_{-6.00546}^{+7.70008}\,(3\sigma)$  \\ \hline
  $l_0$ & \ \ $-1.31401_{-3.56927}^{+3.29455}\,(1\sigma)_{-12.98544}^{+17.06251}\,(2\sigma)\,_{-35.29680}^{+38.88782}\,(3\sigma)$  \ \\
 \hline\hline
 \end{tabular}
 \end{center}
 \caption{\label{tab3} The same as in Table~\ref{tab2}, except
 for SN+CMB data and the prior $0<\beta<1$ only.}
 \end{table}



\vspace{-18mm} 


\section{Concluding remarks}\label{sec4}

The current accelerated expansion of the universe has been one
 of the most important fields in physics and astronomy since
 1998. Many cosmological models have been proposed in the
 literature to explain this mysterious phenomenon. Since the
 nature and cause of the cosmic acceleration are still unknown,
 model-independent approaches to study the evolution of
 the universe are welcome. One of the powerful
 model-independent approaches is the so-called cosmography.
 It only relies on the cosmological principle, without
 postulating any underlying theoretical model. However, there
 are several shortcomings in the usual cosmography. In the
 present work, we try to overcome or at least alleviate these
 problems, and propose two new generalizations of cosmography
 inspired by the Pad\'e approximant. We also confront them with
 the observational data with the help of the Markov chain Monte
 Carlo (MCMC) code {\it emcee}~\cite{ForemanMackey:2012ig},
 and find that they work fairly well.

In the literature (e.g.~\cite{Weinberg2008,Visser:2004bf,
Bamba:2012cp,Dunsby:2015ers,Chiba:1998tc,Neben:2012wc,Cattoen:2007id,
 Cattoen:2008th,Aviles:2012ay,Capozziello:2008tc,
 Vitagliano:2009et,Luongo:2011zz,Busti:2015xqa,Semiz:2015gga}),
 there exist a number of works on cosmography. They focused on
 various issues in cosmology and made much significant
 progress. However, the previous works used
 ordinary cosmography mainly with respect to the redshift $z$
 or the so-called $y$-shift $y=z/(1+z)$. We stress that they are all
 plagued with the problem of divergence (or the unacceptably
 large error when $y\sim 1$), and all fail to predict the
 future evolution of the universe (especially when $z\sim -1$
 or $y<-1$), as mentioned in Sec.~\ref{sec1}. In the present
 work, the two new generalizations of cosmography proposed in
 Secs.~\ref{sec2} and \ref{sec3} can instead avoid or at least
 alleviate these problems of ordinary cosmography. In addition,
 our new generalizations of cosmography are well motivated by
 the Pad\'e approximant, rather than written purely by hand. In
 fact, it has been solidly proved by mathematicians that any
 (even unknown) function can be well approximated by a Pad\'e
 approximant~\cite{r3,r4,r5}, and the Pad\'e approximant often
 gives a better approximation of the function than truncating
 its Taylor series, and it may still work where the Taylor
 series does not converge~\cite{r4}. Noting that ordinary
 cosmography is actually based on a Taylor series, the
 advantages of the Pad\'e approximant mentioned above make our
 new generalizations better than the ordinary cosmography used
 in the literature. Although the Pad\'e approximant has been
 considered in cosmology (see e.g.~\cite{Adachi:2011vu,
 Gruber:2013wua,Wei:2013jya,Liu:2014vda,padeworks}), here we
 instead use it in a fairly different issue. For example, in
 e.g.~\cite{Wei:2013jya,Liu:2014vda} the Pad\'e approximant was
 used to study the issues of an analytical approximation of the
 luminosity distance in the XCDM model, EoS parameterizations,
 and gamma-ray burst cosmology, while we use it to generalize
 cosmography in this work.  By all the above arguments, we
 stress that the present work is significantly different from
 the previous works on both cosmography and the Pad\'e
 approximant in the literature.

The key to avoid or at least alleviate the problems of ordinary
 cosmography is the denominator of the Pad\'e approximant.
 Considering Eqs.~(\ref{eq2}) or (\ref{eq1}), if the order $n$
 of the denominator is larger than or equal to the order
 $m$ of the numerator, this Pad\'e approximant with $n\geq m$
 will not diverge even for $z\gg 1$. In fact, this is just the
 case of our two generalizations (nb. Eqs.~(\ref{eq3}) and
 (\ref{addeq5})). On the other hand, for suitable parameters
 $\beta_i$, it is easy to ensure the denominator not to equal
 zero for the very wide redshift range $-1<z<\infty$, and hence
 the Pad\'e approximant can avoid divergence. The shortcomings
 of ordinary cosmography mainly have their roots in the Taylor
 series. So, generalizing a Taylor series to the Pad\'e approximant
 brings about a possible way out. If all $\beta_i=0$ in the
 denominator of the Pad\'e approximant, it reduces to the usual
 Taylor series. However, a denominator not equal to $1$
 and $0$ makes a big difference.

Here, we would like to clarify the main difference between the
 two new generalizations proposed in Secs.~\ref{sec2} and
 \ref{sec3}. Noting that ordinary cosmography is based on
 a Taylor series, the first one generalizes cosmography by directly
 generalizing the Taylor series to the Pad\'e approximant
 when we expand the luminosity distance $d_L$ (nb. Eq.~(\ref{eq2})).
 On the other hand, the second one instead generalizes
 cosmography by generalizing the redshift $z$ or the $y$-shift
 $y=z/(1+z)$ to the so-called $y_\beta$-shift
 $y_\beta=z/(1+\beta z)$ (which is inspired by the Pad\'e
 approximant), but we still expand the luminosity distance
 $d_L$ in a Taylor series (nb. Eq.~(\ref{eq45})), rather than
 the Pad\'e approximant itself. This is the main difference.
 Noting that $y_\beta=z$ and $z/(1+z)$ when $\beta=0$ and $1$
 respectively, we call $y_\beta=z/(1+\beta z)$
 the ``$y_\beta$-shift'' in analogy to the well-known terminology of
 the ``$y$-shift'' $y=z/(1+z)$ used in the literature, while
 the terminology ``$y$-shift'' comes from the terminology
 ``red-shift'' in fact.

It is of interest to quantitatively compare our two new
 generalizations of cosmography, and also compare them with
 the ordinary cosmography (we thank the referee for pointing
 out this issue). Let us compare our
 $y_\beta$-shift cosmography with the ordinary cosmography
 first. The observational constraints on our $y_\beta$-shift
 cosmography from the SN+CMB data are presented in Table~\ref{tab3}
 and Fig.~\ref{fig5}. It is easy to see that $\beta=0$ and
 $\beta=1$ deviate from the best fit far beyond the $3\sigma$
 confidence level, see especially the leftmost column of
 Fig.~\ref{fig5}. Note that our $y_\beta$-shift reduces to
 the ordinary redshift $z$ and the $y$-shift $y=z/(1+z)$ when
 $\beta=0$ and $1$, respectively. Therefore, our
 $y_\beta$-shift cosmography can fit the SN+CMB
 data significantly better than ordinary cosmography. On the
 other hand, from the last column of Table~\ref{tab1} and
 Table~\ref{tab3}, $\chi^2_{min}=562.171$ and
 $\chi^2_{min}/dof=0.978$ by fitting the Pad\'e parameterization to
 the SN+CMB data, while $\chi^2_{min}=562.654$
 and $\chi^2_{min}/dof=0.979$ by fitting the $y_\beta$-shift
 cosmography to the SN+CMB data. Therefore, our first generalization
 of cosmography is slightly better than the second one.

Other remarks are in order. First, the Pad\'e approximant has
 been used previously in cosmology, e.g. in slow-roll inflation, the
 reconstruction of the scalar field potential, data fitting
 and analytical approximation of the luminosity distance, EoS
 parameterizations, gamma-ray burst cosmology, and cosmological
 perturbations in LSS (see e.g.~\cite{Adachi:2011vu,Gruber:2013wua,
 Wei:2013jya,Liu:2014vda,padeworks}). We advocate further uses
 of the Pad\'e approximant in cosmology. Second, in the present
 work we only derive the generalized cosmography of the
 luminosity distance $d_L$. In fact, one can easily obtain the
 corresponding cosmography of other observable quantities, such
 as the angular diameter distance $d_A$, the photon flux
 distance $d_F$, the photon count distance $d_P$, and the
 deceleration distance $d_Q$, since they can be readily related
 to the luminosity distance $d_L$ (see e.g.~\cite{Cattoen:2008th}).
 Third, we confront the generalized cosmography with the
 observational data only for the spatially flat FRW universe
 ($k=0$). In fact, one can do this for the $k\not=0$ cases
 easily, and hence we do not present them here. Finally, the
 usual cosmography with respect to the redshift $z$ and
 $y$-shift $y=z/(1+z)$ has been extensively applied to
 various cosmological issues, for instance, the EoS of dark energy,
 modified gravity theories like $f(R)$ and $f(T)$ theories,
 gamma-ray burst cosmology, and so on (see
 e.g.~\cite{Weinberg2008,Visser:2004bf,Bamba:2012cp,Dunsby:2015ers,
 Chiba:1998tc,Neben:2012wc,Cattoen:2007id,Cattoen:2008th,
 Aviles:2012ay,Capozziello:2008tc,
 Vitagliano:2009et,Luongo:2011zz,Busti:2015xqa,Semiz:2015gga}).
 The new generalizations of cosmography proposed in the present work
 can also be used in these cosmological issues, and we leave it
 to the future works.


\section*{ACKNOWLEDGEMENTS}
We thank the anonymous referee for quite useful comments and
 suggestions, which helped us to improve this work. We
 are grateful to Zu-Cheng~Chen, Jing~Liu, Xiao-Peng~Yan,
 Shoulong~Li, Hong-Yu~Li, and Dong-Ze~Xue for kind help and
 discussions. This work was supported in part by
 NSFC under Grants No.~11575022 and No.~11175016.

\renewcommand{\baselinestretch}{1.0}


\end{document}